\documentclass[prb,twocolumn,preprintnumbers,superscriptaddress,amsmath,amssymb,floatfix]{revtex4}
\usepackage{graphicx}
\usepackage{dcolumn}
\usepackage{bm}
\begin{document}

\bibliographystyle{apsrev}

\title{Diffusion-limited deposition with dipolar interactions:
fractal dimension and multifractal structure}

\author{M. Tasinkevych}
\affiliation{Max-Planck-Institut f\"{u}r Metallforschung,
             Heisenbergstr. 3, D-70569 Stuttgart, Germany}
\affiliation{Institut f\"{u}r Theoretische und Angewandte Physik, 
         Universit\"{a}t Stuttgart - Pfaffenwalding 57, 
         70569 Stuttgart, Germany}
\author{J. M. Tavares}
\affiliation{Centro de F\'{\i}sica Te\'{o}rica e Computacional 
da Universidade de Lisboa \\
Avenida Professor Gama Pinto 2, P-1649-003 Lisbon, Portugal}
\affiliation{Instituto Superior de Engenharia de Lisboa\\ Rua Conselheiro
Em\'{\i}dio Navarro, 1, P-1949-014 Lisboa, Portugal}

\author{F. de los Santos}
\affiliation{Departamento de Electromagnetismo y F\'\i sica de la Materia,
Universidad de Granada, Fuentenueva, 18071 Granada, Spain}

\date{\today}

\begin{abstract}
Computer simulations are used to generate two-dimensional diffusion-limited
deposits of dipoles.
The structure of these deposits is analyzed by
measuring some global quantities: the density of the deposit and
the lateral correlation function at a given height, the mean height of the upper surface for
a given number of deposited particles and the interfacial width at a given height.
Evidences are given that the fractal dimension of the
deposits remains constant as the deposition proceeds, independently
of the dipolar strength.
These same deposits are used to obtain
the growth probability measure through Monte Carlo techniques.
It is found that the distribution of growth probabilities
obeys multifractal scaling, i.e. it can be analyzed in terms
of its $f(\alpha)$ multifractal spectrum.
For low dipolar strengths, the $f(\alpha)$ spectrum
is similar to that of diffusion-limited aggregation.
Our results suggest that
for increasing dipolar strength both the minimal local growth exponent
$\alpha_{min}$ and the information dimension $D_1$ decrease, while the fractal dimension
remains the same.

\end{abstract}

\pacs{}
\maketitle

\section{Introduction}
The formation of clusters and deposits by irreversible aggregation of particles is an
example of a nonequilibrium  growth process. Although several mechanisms can be involved
in these processes, 
the most simple models only take into account the effects of thermal diffusion. 
Among these models, the DLA (diffusion limited aggregation) \cite{witten_sander:81} and DLD 
(diffusion limited deposition) \cite{meakinbook,meakin:83:2} have been widely used. 
In DLA, particles are released one by one at a distance $R$ from a seed particle, 
and perform a random-walk in a $d$-dimensional space. Eventually,
the random walker either reaches the aggregate and attaches to it, or moves
sufficiently far away from the aggregate to be removed. DLD is a
version of DLA for the growth of deposits on fibers and surfaces.
\cite{meakinbook,meakin:83:2} In DLD, particles also diffuse randomly through a
$d$-dimensional space, but attach either to a $d-1$ dimensional 
substrate or to the deposit.
Despite being based on a very simple algorithm, these models (and modifications thereof) 
generate very complex patterns and have served as a guideline
for the understanding of  wide range of phenomena 
such as electrochemical deposition, viscous fingering, dendritic solidification,
dielectric breakdown, etc. \cite{meakinbook} 

At present, there is no theoretical framework to describe
the scaling behavior of DLA or DLD structures. 
Initially,  it was assumed that
the structure of the aggregates is homogeneous, statistically self-similar
fractal and could be characterized by a fractal dimensionality $D$.
However, this simplified assumption does not fully capture the complexity of DLA and DLD
structures and a better characterization is obtained from the studies
of the growth probability distribution.
One can then apply a multifractal scaling analysis to determine the 
multifractal spectrum of the growth measure, or equivalently an infinite hierarchy of fractal
dimensions. \cite{meakin:85:1,halsey:86:1}
The remarkable scaling behavior of the growth probabilities  \cite{meakinbook}
immediately raises the question of their behavior when mechanisms
other than the thermal diffusion are included.

It is well known that short-range isotropic 
interactions do not change the cluster's structure at large
length scales. At small length scales, however, a short-range attraction (repulsion) 
promotes the formation of less dense (more dense) aggregates but with no change in the 
fractal dimensions $D$. \cite{meakin:83:1,meakin:89:1,vandewalle:95:1}
A different picture emerges when long-range, anisotropic interactions are
considered.  Results for DLA of dipolar particles for $d=2$  indicate that 
$D$ decreases from $D\approx 1.7$ (the value for pure DLA \cite{somfai:03:0})
to $D\approx 1.1$ when the dipolar interaction is increased. \cite{rubis}  This is in accordance with
the results for cluster-cluster aggregation of dipolar particles \cite{mors:87:1} as well as
with experimental results for the aggregation of magnetic micro-spheres. \cite{helgesen:88:1}

\begin{figure}[]
\begin{center}
\includegraphics[width=9cm]{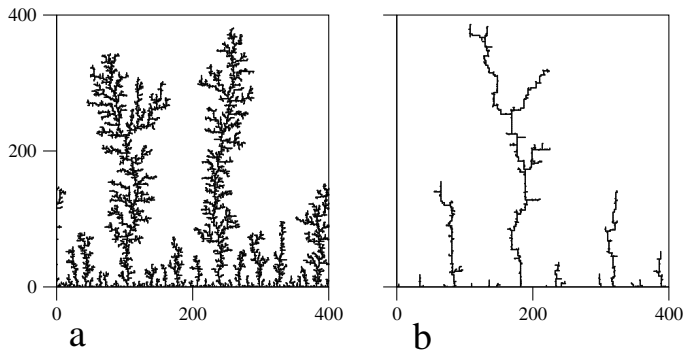}
\end{center}
\caption{Typical DLD deposits which are obtained for a) weak 
and b) strong dipolar interactions. The deposits contain a) 12000 and
b) 2000 particles.
}
\label{two_deposits}
\end{figure}

In our previous work \cite{tasinkevych:04:1} we have grown two-dimensional 
DLD clusters consisting of up to $10^5$ dipolar particles, and
found evidences of a more complex behavior for the deposits' structure. 
Fig.~\ref{two_deposits} shows two typical deposits obtained for weak 
and strong  dipolar interactions. The deposits consist of many tree-like clusters
competing to grow. As the deposition process continues fewer and fewer trees keep on
growing. Eventually only a single tree survives.
In both cases the height and width
of an individual tree of size $s$ (number of particles in the cluster) can be described by 
the power-laws  $H\sim s^{\nu_\parallel}$ and
$W\sim s^{\nu_\bot}$, respectively.  \cite{meakin:86:3} 
Likewise, the average number $n_s$ of trees of a given size scales with $s$ as
 $n_s \sim s^{-\tau}$. \cite{racz:83:0} 
Two scaling regimes have been observed: 
(i) for $s$ less than the crossover size $s^*$, the shape and
the fractal dimension of the trees $D_t$ are temperature dependent in such a way  
that $\tau$ and $D_t$ decrease with increasing of the interaction strength. 
We call this scaling regime the {\em dipolar regime}. 
(ii) For $s> s^*$, and for large enough systems, pure DLD values of the exponents 
are observed independently of the interaction strength. 
This implies a crossover to the diffusion-driven {\em DLA scaling regime}, where 
the effects of the dipolar interactions are dominated by thermal effects. 
The dipolar regime corresponds to deposits exhibiting an orientational order
of dipoles and the onset of the DLA-regime coincides with the disappearance of the
orientational order. \cite{tasinkevych:04:1} 
More strikingly, it  has been found that the value of $D$ (the fractal dimension
of the entire deposit) barely varies with the dipole interaction strength.
In the present paper, we will show further evidence of this by
analyzing in detail the scaling behavior of the density and a lateral correlation function.

Despite having the same fractal dimension, deposits obtained for different 
strengths of the dipolar interaction (equivalently, different effective temperatures) 
exhibit quite different structures. 
This situation resembles that of DLA and
percolating clusters in $d=3$, namely they have the same
fractal dimensions, although their geometrical structures    
are quite different.  \cite{stanley:77:0,schwarzer:92:0}
In this paper we characterize the deposits of dipoles further by 
testing a  multifractal scaling of the growth probabilities $\{p_i\}$, 
where $p_i$ is the probability that the perimeter site $i$ is the 
next to be added to a cluster.

This paper is organized as follows: in section \ref{sec:model} we
introduce the model of dipolar DLD and describe  details of   
the simulations. In section \ref{subsec:paco}, the scaling exponents   
obtained for the density, lateral correlation function,
the mean height of the upper surface and the interfacial 
width are compared with those of DLD. 
In section \ref{subsec:mult} we present the multifractal 
spectra calculated for strong and weak dipolar interactions, 
and for several stages of growth. 
Finally, in section \ref{sec:conclusion} we summarize our findings.

\section{Model and Simulations Details}
\label{sec:model}

The model and the simulation technique are the same as in our previous 
works \cite{tasinkevych:04:1, tasinkevych:03:1}; we briefly outline them here, and 
describe simulation details which allowed us to grow relatively large clusters. 
The simulations are performed on a two-dimensional square-lattice 
with lattice spacing $a$ and width $La$.
The adsorbing substrate corresponds to the bottom row ($y=0$) and
periodic boundary conditions are applied in the direction parallel 
to the substrate ($x$-direction). 
Particles carry a three-dimensional dipole moment of strength $\mu$ and interact with other 
particles through the dipolar pair-potential $\phi_D$:
\begin{eqnarray}
\label{dipol_inter1}
\phi_D(1,2) = -\frac{\mu^2}{r^3_{12}}
\left[ 3(\hat{\mbox{\boldmath $\mu$}}_1 \cdot \hat{\mbox{\boldmath $r$}}_{12}) 
({\hat{\mbox{\boldmath $\mu$}}_2} \cdot  \hat{\mbox{\boldmath $r$}}_{12}) 
-{\hat{\mbox{\boldmath $\mu$}}_1} \cdot {\hat{\mbox{\boldmath $\mu$}}_2} \right],  
\end{eqnarray}
where $r_{12}=|\vec{\mbox{\boldmath $r$}}_1 - \vec{\mbox{\boldmath $r$}}_2| \geq a$, 
$\hat{\mbox{\boldmath $r$}}_{12} = (\vec{\mbox{\boldmath $r$}}_1 - \vec{\mbox{\boldmath $r$}}_2)/r_{12}$,
$\vec{\mbox{\boldmath $r$}}_i$ is a two-dimensional vector and
 ${\hat{\mbox{\boldmath $\mu$}}_1}$,
${\hat{\mbox{\boldmath $\mu$}}_2}$ are three-dimensional unit vectors in the 
direction of the dipole moments of particles 1 and 2 respectively. 

Initially, a particle with  random dipolar moment is introduced at the lattice site $(x_{in},H_{max}+A\,L)$,
where $x_{in}$ is a random integer in the interval $[1,L]$, $H_{max}$
is the maximum height of the deposit and $A$ is a constant. The particle then 
diffuses by a series of jumps to nearest-neighbor lattice sites, while
interacting with the particles that are already part of the deposit. 
Eventually, the particle either contacts
the deposit (i.e., becomes a nearest-neighbor of another particle that belongs to the deposit), 
or attaches to the substrate (i.e., reaches the
bottom of the simulation box), or moves away from the substrate.
In the latter case, the particle is removed when it reaches a distance from the substrate
larger than $H_{max}+2A\,L$, and a new one is introduced.
Once the particle attaches to the substrate or to the deposit, its dipole 
relaxes along the direction of the local field created by all other
particles in the deposit.
In all simulations reported here we take $A=1$; larger values of $A$ 
have also been tested and found to give the same results, but with drastically 
increased computational times.

The effect of the dipolar interaction on the random walk is incorporated 
through a Metropolis algorithm.
The interaction energy of an incoming particle with the deposit is given by 
$E(M,\vec{\mbox{\boldmath $r$}},{\hat{\mbox{\boldmath $\mu$}}})=\sum_{i=1}^M
\phi_D(i,M+1)$, where $\vec{\mbox{\boldmath $r$}}$ denotes the particle's position,
${\hat{\mbox{\boldmath $\mu$}}}$ is the orientation of its dipole moment and 
$M$ is the number of particles in the deposit.
To simplify the notation, the sum over the periodic replicas
of the system is omitted in the last expression. 
To move the particle, we select a neighboring site
$\vec{\mbox{\boldmath $r$}}^{\prime}$ and a new dipole orientation 
${\hat{\mbox{\boldmath $\mu$}}}^{\prime}$ randomly. 
This displacement is accepted with probability
\begin{equation}
\label{metropolis}
p={\rm min}\left\{1,\exp\left[- \frac{E(M,\vec{\mbox{\boldmath $r$}}^{\prime},
{\hat{\mbox{\boldmath $\mu$}}^{\prime}})
-E(M,\vec{\mbox{\boldmath $r$}},\hat{\mbox{\boldmath $\mu$}})}{T^*} \right]\right\},
\end{equation}
where $T^*=k_BTa^3/\mu^2$ is an effective temperature 
inversely proportional to  dipolar energy scale. 
In the limit $T^*\to 0$, only displacements which lower the energy $E$ are accepted. On the other extreme,
for $T^*\rightarrow \infty$, all displacements are accepted and our model reduces to DLD.

One can estimate the number of calculations of $\phi_D$,  $t$,  necessary to
grow a deposit of mass $M$ on a strip of width $L$ in the following way. Since each incoming particle
starts its movement at a distance of order $L$ from the deposit, 
it takes roughly $L^2$ steps to reach the deposit. If there are $n$ particles
in the deposit, $\phi_D$ has to be calculated $n L^2$ times. Summing this
factor over $M$ particles gives $t \sim L^2 M^2$. 
Since we have performed simulations for $L\sim 10^3$ and 
$M\sim 10^5$, these values roughly give $t \sim 10^{16}$, which is about 
3 to 4 orders of magnitude larger than the corresponding values of the 
largest systems simulated in an equilibrium run of two-dimensional dipolar fluids. \cite{tavares:02:1}
Therefore, to make simulations feasible we rewrite the expression for the 
incoming particle--deposit interaction energy in a different form. 
To this aim, the dipolar pair potential $\phi_D$, Eq.~(\ref{dipol_inter1}), 
is rewritten as  $\phi_D(i,j) = \mu \vec{\mbox{\boldmath$D$}}_{ij}\cdot \hat{\mbox{\boldmath $\mu$}}_j$, 
where    
$\vec{\mbox{\boldmath$D$}}_{ij}$ is the dipolar field created by the particle
$i$ at the site occupied by the particle $j$
\begin{equation}
\label{dipol_inter2}
\vec{\mbox{\boldmath $D$}}_{ij}= -\frac{\mu}{r^3_{ij}}
\left[ 3(\hat{\mbox{\boldmath$\mu$}}_i \cdot \hat{\mbox{\boldmath $r$}}_{ij})  
\hat{\mbox{\boldmath $r$}}_{ij} -\hat{\mbox{\boldmath $\mu$}}_i \right],
\end{equation}
Using Eq.~(\ref{dipol_inter2}) one can write the interaction energy between the
incoming particle (which is at site $\vec{\mbox{\boldmath $r$}}_j$  and has
a dipole orientation $\hat{\mbox{\boldmath $\mu$}}_j$) and a deposit formed by $M$
dipoles as
\begin{equation}
\label{dipol_field}
E(M,\vec{\mbox{\boldmath $r$}}_j,\hat{\mbox{\boldmath $\mu$}}_j)=\mu
 \vec{\mbox{\boldmath$U$}}(M,j)\cdot \hat{\mbox{\boldmath $\mu$}}_j, 
\end{equation}
where $\vec{\mbox{\boldmath$U$}}(M,j) \equiv \sum_{i=1}^M \vec{\mbox{\boldmath$D$}}_{ij}$ 
(again, the sum  over periodic replicas is omitted) is the dipolar field
created by the deposit at site $\vec{\mbox{\boldmath $r$}}_j$. 
During  simulations we store  
the dipolar field $\vec{\mbox{\boldmath$U$}}(j)$ at site $\vec{\mbox{\boldmath $r$}}_j$  
together with the size of the deposit $M(j)$ for which the field $\vec{\mbox{\boldmath$U$}}$ is calculated.
The particles in the deposit are ordered according to their arrival ``time''.  
Then, the interaction energy is rewritten in the form
$E(M,\vec{\mbox{\boldmath $r$}}_j,\hat{\mbox{\boldmath $\mu$}}_j)
 = \mu (\vec{\mbox{\boldmath$U$}}(j) +\sum_{i=M(j)+1}^M\vec{\mbox{\boldmath$D$}}_{ij} )\cdot
\hat{\mbox{\boldmath $\mu$}}_j$,  which allows to reduce the number of calculations of the 
dipolar pair-potential from $M$ to $M-M(j)$.
At the early stages of growth most of the lattice sites have not yet been visited 
by an incoming particle, hence  $M(j)=0$.  However, as the deposit grows 
the number of the lattice sites that have been visited more than once
increases. Moreover, due to the shadowing effect, the sites close to the 
tips of the trees become visited more frequently, and thus
the overall computation time of the dipolar energies is decreased.  
We have performed tests to compare the computation times 
for calculating the energy between the two methods. For a system of width 
$L\sim 10^3$ and $M\sim 10^3$ particles, the storage of the dipolar fields represents
gains in computational time of about 3 orders of magnitude. 

Finally, the long-range of the dipolar interaction is treated by the Ewald sum method 
adapted to the slab geometry of the system. \cite{tasinkevych:03:1}

\section{Results}
\label{sec:results}

Simulations were carried out using four temperatures,
$T^*=10^{-1}, 10^{-2}, 10^{-3}, 10^{-4}$,
and  four system sizes, $L=200, 400 , 800, 1600$ with
20 000, 30 000, 50 000, and  100 000 particles per deposit,
respectively. Each choice of these parameters corresponds to
one of the two regimes of growth. For instance, the DLA regime is never
observed for the lowest temperature $T^*=10^{-4}$, while for $T^*=10^{-1}$ it is easily reached
even for the smallest system size $L=200$. As we mentioned above, our previous findings suggest 
that the fractal dimension of the deposits $D$ is the same as for DLD even in 
the dipolar regime. \cite{tasinkevych:04:1} To validate this picture, we next 
examine the scaling behavior of the following quantities: the density of the deposit and 
the lateral correlation function at a given height; the mean height of the upper surface for
a given number of particles; the interfacial width at a given height. 
All of them are global quantities that characterize the entire deposit, and which 
do not show any qualitative or quantitative differences between the dipolar and the 
DLA regimes.   

\subsection{The fractal dimension}
\label{subsec:paco}

\begin{figure}[]
\begin{center}
\includegraphics[width=7cm]{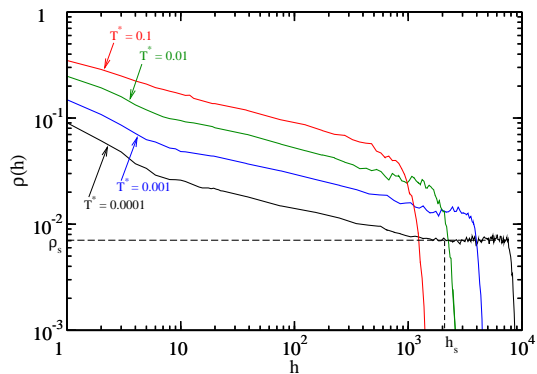}
\end{center}
\caption{Mean density $\rho$ of deposits as a function of the distance $h$ from the substrate.
The densities are calculated for a system-size $L=1600$ and several values of the effective 
temperature $T^*$. Beyond the crossover height $h_s$, the density saturates to a constant value 
$\rho_s$. The  horizontal and vertical dashed-lines indicate the $\rho_s$ and $h_s$ obtained 
for $T^*=10^{-4}$. }
\label{densities}
\end{figure}

A typical plot of the mean density
$\rho(h)$ at a distance $h$
from the substrate has three regimes separated by two
crossover heights, $h_i$ and $h_s$ (see Fig.~\ref{densities}).
At early ``times'' the deposit builds up until it reaches
a height $h_i$. Then, there appears a scaling
regime during which the density decreases as a power
of height, $\rho(h) \sim h^{-\alpha}$,
the exponent $\alpha$ being related to the fractal dimension
$D$ of the deposit by $D=2-\alpha$. \cite{meakin:84}
The density stops decreasing and saturates when the lateral
correlation length $\xi_\parallel$ reaches the size of the
system $L$. Given that $\xi_\parallel$ grows with height as
$\xi_\parallel \sim h^\zeta$, then $h_s \sim L^\gamma$, with
$\gamma = \zeta^{-1}$. This behavior can be described by the
scaling form
\begin{equation}
 \rho(h,L)\sim L^{-\beta} f(h/L^\gamma),
\label{scaling_dens}
\end{equation}
where $f(x)$ is a scaling function with the properties:
$f(x) \sim x^{-\alpha}$ for small $x$ and $f(x) \to const$
for large $x$. It is clear that $\beta=\alpha \gamma$.
A linear regression between $h_i$ and $h_s$
for the largest system-size ($L=1600$)
gives roughly the same scaling exponent
$0.25< \alpha <0.29$, showing no systematic variation with temperature. This
implies a value of the fractal dimension $1.71< D <1.75$, which is the same
as for DLD. As we mentioned above, this is in contrast with a previous study, 
Ref.~\onlinecite{rubis}, where a continuous variation of $D$ as a function of $T^*$ was reported.
Although several source of discrepancy are possible, \cite{differences}
we believe that the estimates of Ref.~\onlinecite{rubis}
suffer from insufficient statistics and rather small cluster masses, that
prevent the true asymptotic regime ($M\to \infty$) from being reached.

\begin{figure}[]
\begin{center}
\vspace*{0.2cm}

\includegraphics[width=7cm]{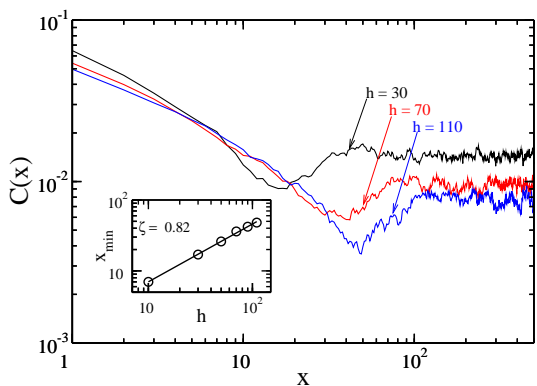}
\end{center}
\caption{Lateral density-density correlation function calculated according to Eq.~(\ref{eq:corrs}) for several
distances $h$ from the substrate. The system-size is $L=1600$ and the effective temperature  $T^*=0.1$.
The position of the minimum $x_{min}$ of the correlation function as a function of $h$ is shown in the inset.
$x_{min}$ scales with $h$ with a scaling exponent $\zeta \approx 0.82$. 
}
\label{corrs}
\end{figure}

According to Eq.~(\ref{scaling_dens}), for $h \gg L$ the mean density
attend a constant value $\rho_s$ that scales with the system-size as
$\rho_s \sim L^{-\beta}$. For DLD the exponent $\beta \approx
0.33$. \cite{meakinbook} This value agrees with those
obtained for $T^*=10^{-4}$ and $T^*=10^{-3}$ (the only temperatures for
which density saturation is neatly observed), $\beta= 0.33(1)$ and $\beta =0.34(1)$, 
respectively. From the scaling relation $\gamma = \beta / \alpha$
one has $\gamma =1.3(1)$.  An independent check of this value
was carried out by calculating the height $h_s$ at which the density saturates 
and which is related to the system-size by $h_s \sim L^\gamma$.
There is substantial uncertainty in the determination of $h_s$
and, again, we are limited to the two lowest temperatures, $T^*=10^{-4}$ and $10^{-3}$.
For both of them $\gamma =1.30(5)$, while for DLD a value of
$\gamma \approx 1.20$ has been measured.

There is a third route to the exponent $\gamma$ through the two-point lateral density-density
correlation function $C(x,h)$ for horizontal cuts at a height $h$. This correlation
function is defined as
\begin{equation}
C(x,h)=\frac{1}{L} \sum_{x'} \sigma(x'+x,h) \sigma(x',h),
\label{eq:corrs}
\end{equation}
where $\sigma(x,h)$ is an occupation number which equals unity if the lattice site $(x,h)$
belongs to the deposit and zero otherwise.
Fig.~\ref{corrs} depicts a plot of $C(x,h)$ as a function of $x$
for several  heights $h$. The main feature
is a pronounced minimum at $x_{min} \approx h^\zeta$  which
can be interpreted as the mean distance between the trees  at the
height $h$, Ref.~\onlinecite{meakin:86:3}, and that, for DLD, scales with $h$
with  an exponent $\zeta \approx 0.80 - 0.85$.
For very large heights, the exponent $\zeta$  increases slowly with increasing height
and may approach unity.\cite{meakin:86:3}
Owing to poor statistics, our simulation data
enables us to evaluate $\zeta$ only for $T^*=10^{-1}$ and $10^{-2}$,
thereby extending our estimations of $\gamma$ to the entire range
of available temperatures. Our results are
consistent with $\zeta \approx
0.85$, equivalently $\gamma \approx 1.2$, and no
significant deviation from DLD behavior is found.

Some other aspects of the morphology of the
deposits have also been investigated. In our
earlier work, \cite{tasinkevych:03:1} the mean height $\overline{h}$
of the upper surface  $h(x)$ ($h(x)$ is defined as the maximum height of the occupied
sites which are in the column $x$)  was shown to scale
with the number of particles $M$ (in the large-$M$
limit and for every temperature) with an effective exponent
$\phi = 1/(D-d+1) \approx 1.33 - 1.44$, which corresponds
to a fractal dimension $D \approx 1.69 - 1.75$.  In the present paper 
we double the
number of particles of the deposits to $10^5$, which allows an
improvement of these bounds to $D \approx 1.71 -1.73$. As we wrote in Ref.~\onlinecite{tasinkevych:03:1},
 had the deposits been allowed to grow only to
intermediate stages (e.g., $10^4$ particles) an apparent
variation of $D$ with $T^*$ would have obtained.

Owing to scale invariance, the exponent $\phi$ should coincide
with the {\it growth} exponent which controls
the power-law  divergence  of the  width of the upper surface
$W=\langle[h(x) -\overline{h}]^2 \rangle^{1/2}\sim M^\phi$.
Linear fits yield $\phi \approx 1.3 - 1.4$,
in agreement both with our previous
estimation and with the value we have measured for DLD,
$\phi = 1.34(2)$.
The divergence of the width is cut off at the the system-size length scale, and the
width reaches a saturation value $W_{sat} \sim L^\delta$, where
the exponent $\delta$ is a conventional measure of
interface roughness. Unfortunately, for every temperature
saturation is barely reached
only for $L=200$, whilst for the rest of sizes much
bigger deposits would be necessary.

\subsection{Multifractal spectrum in the dipolar and the DLA scaling regimes}
\label{subsec:mult}

We discuss hereafter the dependence of the multifractal spectrum of the growth-measure on 
the system size, the total number of particles, and the effective temperature.

\subsubsection{General remarks}

\begin{figure}[]
\begin{center}
\includegraphics[width=9cm]{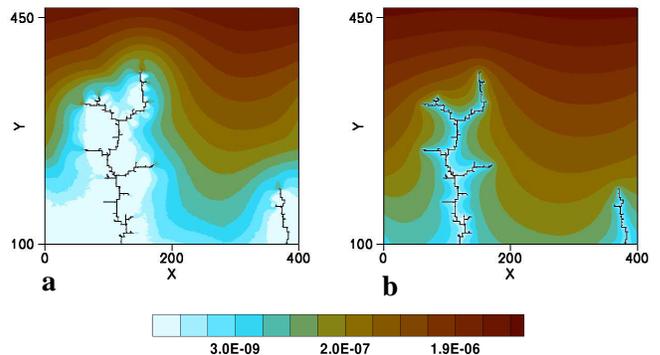}
\end{center}
\caption{
Iso-lines of the visit probability distribution in the vicinity of a
deposit of dipoles. The probabilities are estimated numerically by using probe particles and 
counting the frequency of visiting of each unoccupied lattice site.
The deposit contains $2000$ dipoles and has been grown at $T^*=10^{-4}$.
In a) the strength of the dipolar interaction of the probe particles with the 
dipoles in the deposit corresponds to an effective temperature $T^*=10^{-4}$. In b) the
interaction is turned off ($T^*=\infty$) and the probe particles follow random-walk
trajectories. Dark and light regions correspond to
high and low visit probabilities, respectively. 
}
\label{effects_of_dipoles}
\end{figure}

For DLA and related models, the growth probability varies sharply from site to site,
changing from large values at the outer tips to very low values for the less
accessible points deep within the cluster. It has been found that long-range 
repulsions enhance the screening effects,
which leads to the formation of less branched clusters. Contrarily,
long-range attractions reduce the growth probability at the
cluster's tips and lead to the formation of more compact objects.
In Fig.~\ref{effects_of_dipoles}.a we show  the iso-lines of the visit 
probability distribution in the vicinity of a deposit grown at $T^*=10^{-4}$.
To estimate the visit probabilities  we use probe particles that follow biased random-walk
trajectories, and register the frequency of visiting the unoccupied lattice sites. 
To illustrate the effects of the dipolar interaction, in Fig.~\ref{effects_of_dipoles}.b we show 
the visit probabilities around the same deposit estimated under purely diffusive conditions ($T^*=\infty$),
i.e. the dipolar interaction between the probe particles and the particles in the deposit 
has been turned off. The interaction strongly enhances screening behavior of
the cluster. More perimeter sites become screened, i.e. they are characterized by  smaller
values of the sticking probability.
On the other hand, the dipolar interaction enhances  the  concentration of the
growth-measure at the outer sites of the cluster's branches. 

Studies of the growth probability distribution for DLA 
\cite{ball:90:0,amitrano:86,hayakawa:87} and related models \cite{meakin:87:1}
indicate that the distribution of the growth-measure can be described in
terms of a multifractal scaling model. Suppose that the boundary of a cluster 
of size $L$ is covered by $S(\epsilon)$ boxes of size $\epsilon$. 
Then, a growth measure $p(i)$ is introduced in the $i$'th box as the probability for 
a random-walker to land in box $i$. 
As mentioned above, $p(i)$ varies sharply over the cluster surface in such a way that the local 
growth probability density diverges at the tips and goes to 
zero inside the ``fjords'' (both behaviors are cut off at the particle length scale). 
To each box one can associate a singularity exponent $\alpha(i)$ via
\begin{equation} 
p(i) \sim \Bigl (\frac{L}{\epsilon}\Bigr )^{-\alpha(i)}.
\label{eq:p-scaling}
\end{equation}
In what follows we keep $\epsilon$ fixed at the lower cut-off length scale,
which is of one lattice unit $\epsilon \equiv 1$, and $L/\epsilon$
is varied by examining systems with different sizes $L$.
The behavior of the number of boxes $N(\alpha) d\alpha$ with the scaling exponent
$\alpha$
taking on a value in the interval $[\alpha,\alpha+d\alpha]$
defines the local scaling density exponent $f(\alpha)$
\begin{equation}
N(\alpha) \sim  L^{f(\alpha)}.
\label{eq:Nalpha-scaling}
\end{equation}
One then introduces the scaling function $\tau(q)$ which characterizes the scaling behavior 
of the moments of the probability measure
\begin{equation}
Z(q)\equiv \sum_i p(i)^q \sim L^{-\tau(q)}.
\label{eq:moments}
\end{equation}
Using Eqs.~(\ref{eq:p-scaling}) and (\ref{eq:Nalpha-scaling}), $Z(q)$ can be written in the form
\begin{eqnarray}
Z(q) \sim  \int d\alpha N(\alpha)p^q =
 \int d\alpha L^{f(\alpha) - q\alpha }.
\label{eq:moment-integral}
\end{eqnarray}
Evaluating this integral by saddle-point approximation leads to
\begin{equation}
Z(q) \sim  L^{f(\alpha(q)) - q\alpha(q)},
\label{eq:t-q-alpha-f}
\end{equation}
where the functions $\alpha(q)$ and $f(\alpha(q))$ are defined implicitly by the condition
\begin{equation}
\frac{df(\alpha)}{d\alpha} = q.
\label{eq:alpha-q}
\end{equation}
By comparing  Eq.~(\ref{eq:t-q-alpha-f}) with Eq.~(\ref{eq:moments})  the following relation is obtained
\begin{equation}
-\tau(q) = f(\alpha(q)) - q\alpha(q).
\label{eq:assymptotic_t-f-alpha_relation}
\end{equation}
The moment scaling function $\tau(q)$ is related to the family of dimensionalities
$D_q$ introduced by Hentschel and Procaccia \cite{hp} through $\tau(q)=(q-1)D_q$. 
The limit $D_0 \equiv D \equiv \lim_{q \rightarrow 0^+}D_q$ is the fractal dimension
of the cluster, and $D_1 \equiv \lim_{q \rightarrow 1^+}D_q$ is the information dimension.
For DLA, $\tau(q)$ is a non-linear function of $q$, i.e. an infinite hierarchy of exponents 
is required to characterize the moments of the probability  measure.
Some points of the $f(\alpha)$ spectrum can be related directly to the $D_q$: 
the maximum of $f$ is given by $f(\alpha(q=0))=D$,
$f(\alpha(q=1))=\alpha=D_1$, $\alpha_{min}=D_{\infty}$, and $\alpha_{max}=D_{-\infty}$.

Usually, $\tau$ is calculated as a function of $q$ and then 
the multifractal spectrum $f(\alpha)$ is obtained after performing a Legendre transform. 
The multifractal spectra that we present here, however,
are obtained through interpolation of numerical histograms, using equations (\ref{eq:p-scaling}) and (\ref{eq:Nalpha-scaling}). 
Because of the unknown normalization constants in these scaling
relations, this method provides local growth exponent $\alpha$ and local scaling density $f(\alpha)$ up to 
additive corrections of order $\sim 1/\ln L$, what causes the $f$ spectra to fail to 
satisfy a number of important properties including data collapse, tangency of the $f=\alpha$ line
and the $f(\alpha)$ curve at a single point (corresponding to $q=1$ and giving $D_1$), or mislocation of
other representative points. Here, we shall show that taking into account 
these correction terms improves considerably the measured spectra (at least for low $\alpha$).

With this purpose, first note that for a finite $L$ all quantities $p,N,\alpha,f,\tau$ must carry 
a label $L$. We define the size dependent $\alpha_L$ and $f_L$ as following:
$p_L \equiv L^{-\alpha_L}$, $N_L(\alpha_L) \equiv  L^{f_L(\alpha_L)}$.  
To bring in the corrections to scaling, the integral (\ref{eq:moment-integral}) 
is evaluated retaining the next-order terms to give
\begin{equation}
Z_L(q) \equiv L^{-\tau_L(q)} \approx L^{f_L(\alpha_L) - q\alpha_L(q)-\frac{\ln(-f_L^{\prime\prime})}{2\ln L}-G(L)},
\label{eq:finite_size_t-q-alpha-f}
\end{equation}
where $G(L) = \ln[\ln L/(2\pi)]/(2\ln L)$ and $f_L^{\prime\prime}$ stands for the second 
derivative of $f_L$ with respect to $\alpha_L$. Notice that 
\begin{equation}
\frac{df_L}{d\alpha_L}=q
\label{eq:finite_size_alpha-q}
\end{equation} 
holds. Assuming now that, to order $O(1/\ln L)$,  $\tau_L \approx \tau$ and using
Eqs.~(\ref{eq:finite_size_t-q-alpha-f}),(\ref{eq:finite_size_alpha-q}) the following expressions can be readily derived
\begin{equation}
\alpha(q) = \alpha_L(q) + \frac{f_L^{\prime\prime\prime}}{2(f_L^{\prime\prime})^2\ln L},
\label{eq:alpha_assympt}
\end{equation}

\begin{eqnarray}
f(\alpha) = f_L(\alpha_L)+\frac{f_L^{\prime}f_L^{\prime\prime\prime}}{2(f_L^{\prime\prime})^2\ln L} 
-\frac{\ln(-f_L^{\prime\prime})}{2\ln L}-G(L).
\label{eq:f_assympt}
\end{eqnarray}
Taking $\alpha_L$ as the independent variable, instead of $q$, then these two equations give the
parametric representation of the asymptotic multifractal spectrum $f(\alpha)$ as a function of 
measured quantities $f_L,\alpha_L$.

\subsubsection{Calculation of $f_L(\alpha_L)$ trough numerical histograms}

The growth probabilities $p_L(i)$ are estimated numerically using probe particles that sample the growth
measure. The probe particles follow biased random-walk trajectories in the dipolar field created by
the cluster. Then $p(i)$  is calculated as $p_L(i) = N_L(i)/N_{T}$, where  $N_L(i)$ is the number of trajectories 
which have terminated on the perimeter site $i$ and $N_T$ is the total number of trajectories (probe particles).
Having estimated the growth probabilities $p_L(i)$, 
we then calculate the number of perimeter sites $N_L(\alpha_L)\Delta\alpha_L$ with the value of $p_L(i)$ in the interval
$[L^{-\alpha_L}, L^{-\alpha_L+\Delta\alpha_L}]$, where $\alpha_L$ is defined as 
\begin{equation}
\alpha_L = -\frac{\ln p_L(i) }{\ln L}.
\label{eq:p-L_simpl}
\end{equation}
More specifically, we calculate the histogram $N_L(\alpha_L)$ from which the 
multifractal spectrum is obtained as 
\begin{equation}
f_L(\alpha_L) = \frac{\ln N_L(\alpha_L)}{\ln L}.
\label{eq:N_alpha-L_simpl}
\end{equation}
In the calculations we choose for the width of the bin $\Delta\alpha_L\approx0.15$.

\begin{figure}[]
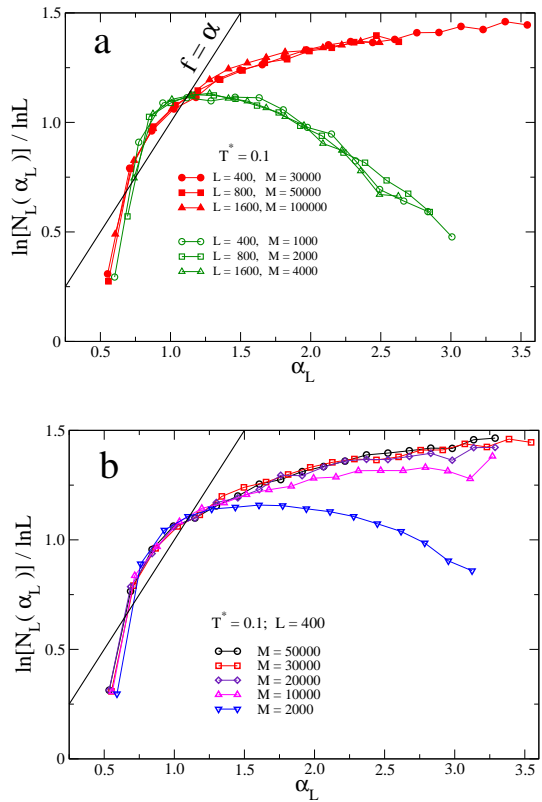

\begin{center}
\includegraphics[width=7cm]{figures/T_0.1a.eps}
\vspace*{0.5cm}

\includegraphics[width=7cm]{figures/T_0.1b.eps}
\end{center}
\caption{ $f_L(\alpha_L)$ spectrum of the growth probability measure for DLD of dipolar particles calculated
at an effective temperature $T^*=0.1$ as defined in section~\ref{sec:model}.
This value of $T^*$ corresponds to the DLA scaling regime for all values of $M$ presented in the figure.
$\alpha_L = -\ln(p_L)/\ln L$ with $p_L$ being the growth measure.
(a) Results obtained for system-sizes $L=400,800,800$, and deposit masses in the range
$10^3\leq M \leq 10^5$. (b) demonstrates the convergence of $f_L(\alpha_L)$ spectrum
towards an asymptotic regime for $M>2\times 10^4$. We find a minimal value of the local growth exponent $\alpha_{Lmin}\approx 0.56$. 
The solid straight lines are given by $f_L=\alpha_L$.} 
\label{fig:mult1}
\end{figure}

\begin{figure}[]
\begin{center}
\includegraphics[width=7cm]{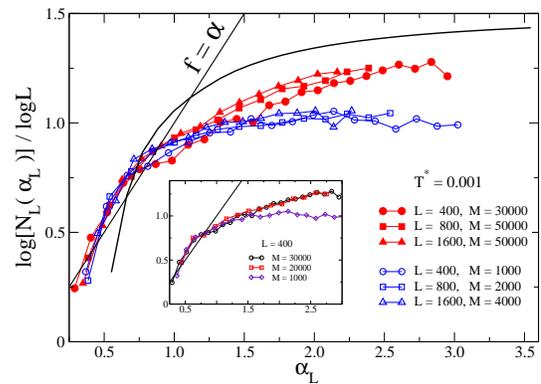}
\end{center}
\caption{ $f_L(\alpha_L)$ spectrum of the growth probability measure for DLD of dipolar particles calculated
 at $T^*=0.001$, with $\alpha_L = -\ln(p_L)/\ln L$.
 The results are obtained for system sizes $L=400,800,1600$, and deposit masses in the range
$10^3\leq M \leq 5\times 10^4$. The solid curve schematically represents the $f_L(\alpha_L)$ spectrum for $T^* = 0.1$.
 The inset demonstrates the convergence of the $f_L(\alpha_L)$ spectrum towards an asymptotic regime for $M>2\times 10^4$.
 The values of  deposit masses $M\leq 4000$ correspond to the dipolar scaling regime;  the individual
 trees of the deposits have the fractal dimension $D_t \approx 1.33$. 
 For  deposit masses $2\times 10^4 \leq M \leq 5\times 10^4 $ the system is
 in the DLA scaling regime with $D_t \approx 1.56$.
  We observe the minimal value of the local growth exponent $\alpha_{Lmin}\approx 0.36$ in both dipolar
 and DLA scaling regimes.}
\label{fig:mult2}
\end{figure}

High effective-temperatures results are shown in 
Fig.~\ref{fig:mult1}. $f_L(\alpha_L)$ is obtained at $T^*=0.1$ for deposits of size 
$L = 400,800,1600$ and number of particles $M$  in the range
$10^3 \leq M \leq 10^5$. 
The number of probe particles used varies from $10^8$ for $(L,M)=(1600,10^5)$ 
to $1.6\times10^9$ for $(L,M)=(400,3\times10^4)$.
The collapse of the curves shows evidence that the distribution of the
measure is multifractal,
with $f_L(\alpha_L)$ gradually converging towards a well-defined spectrum 
as the deposit mass increases. 
An illustration that the uppermost curves are asymptotic is shown in 
Fig.~\ref{fig:mult1}b which displays data only for deposits of size $L=400$ and containing  
$M=$ 2000, 10000, 20000, 30000 and 50000 particles (notice the small change in $f_L$ when
$M$ changes from 30000 to 50000).
 The minimal value of the local
growth exponent $\alpha_{L}$ is associated with the region where the measure is  most
concentrated. It is found that $\alpha_{Lmin} \approx 0.56$,  a value that compares favorably with measurements 
for DLA for $d=2$, $\alpha_{min}\approx 0.59(4)$. 
\cite{ball:90:0}
$\alpha_{min}$ is related to the fractal dimension of the deposit $D$ through $D \geq 1+\alpha_{min}$. \cite{halsey:book}
This inequality is in agreement with the values of $D$ estimated by the scaling of the densities. 
One can identify a region of small $\alpha_L$'s for  which
the  multifractal spectrum does not change with the deposit mass $M$.
In particular, the value of $\alpha_{Lmin} \approx 0.56$ is the same for 
the biggest and the smallest deposits, ($L=1600,M=10^5$) and ($L=400, M=10^3$), 
what suggests that the maximum growth probability, within the considered range of $M$, does not depend on $M$. 
Beyond the boundary value $\alpha_L \approx 1.1$ there is  clear mass dependence.
Finally, notice that $f_L(\alpha_L)$ flattens for larger values of $\alpha_L$.
The maximum possible value of $f_L$, $f_{max}$, should  be equal the fractal dimension $D$ of the deposit. 
For the DLA the maximum has been reported to be at the position $\alpha_0 \approx 4.5$, Ref.~\onlinecite{ball:90:0}. 
With the Monte-Carlo technique we can not determine such small probabilities. However,
one can anticipate that $f_{max}$ is not far from the expected value $D$. 
 
Low effective temperatures results are displayed in
Fig.~\ref{fig:mult2}. $f_L(\alpha_L)$ is calculated at $T^*=10^{-3}$ for both
the initial and late stages of growth (small and large $M$),
the number of probe particles varying from $7.9\times10^6$ for $(L,M)=(1600,5\times10^4)$ 
to $5\times10^7$ for $(L,M)=(400,3\times10^4)$.
The initial stages of growth correspond to the dipolar regime, 
with the deposits containing $M=1000, 2000, 4000$ particles 
for $L=400, 800, 1600$, respectively. 
At the late stages of growth ($M=30000, 50000, 50000$ for  $L=400,800,1600$ respectively),
the DLA regime has been attained, as described in section IIIA. 
The inset of Fig.~\ref{fig:mult2} demonstrates the convergence of $f_L(\alpha_L)$ towards 
the asymptotic behavior which is reached at $M>20000$. 
Again, like in the case  $T^*=0.1$, a region of small $\alpha_L$'s can be identified
where the multifractal spectra does not depend on $M$.
In fact, in both regimes we obtain a value of $\alpha_{Lmin} \approx 0.36$,
which is considerably lower than the corresponding value obtained for $T^*=0.1$.
The turning point separating $M$-dependent from $M$-independent behavior is located at
$\alpha_L \approx 1.35$. 
Beyond this point good data collapse is still observed for the dipolar regime, 
whereas the collapse of the curves is not impressive after the crossover to the DLA regime
has taken place. We shall see below how to amend this shortcoming. 
The changes in the shape of the $f_L$-spectrum as $M$ grows  can be interpreted as an increase 
in the fractal dimension of the screened parts of the deposit,  but for the 
studied range of $\alpha_L$'s, $f_L$ still stays well below the $T^*=0.1$ curve. 
According to the picture developed in section \ref{subsec:paco}, 
the maximum values of $f_L$ should equal $D\approx 1.71$. The
figure suggests a further increase in $f_L$ if more probe particles are used 
to estimate the growth probabilities. Although no conclusive 
statements can be made regarding this point, we shall see in the next subsection 
that it is expected that the large $\alpha_L$ parts of the spectra shift upwards once corrections to scaling 
have been included. 
 
Lastly, the elongated appearance of the trees in the dipolar regime 
resembles that observed in the initial stages of deposits grown at
high temperatures in the DLA regime. Our data show, however, 
that the multifractal structure of the dipolar regime is clearly different
from those of the initial stages of growth of deposits at higher temperatures, 
even if the fractal dimensions are similar.

\subsubsection{Corrected spectra}

When the multifractal measure possesses a continuous spectrum, the straight line $y=\alpha$ is tangent 
to the $f(\alpha)$ curve at $f(\alpha)=\alpha$. \cite{halsey:88:0} 
This general behavior is not seen in the hitherto shown spectra because they have 
been obtained through numerical histograms. Hence, the scaling
relations expressed by  Eqs.~(\ref{eq:p-scaling}), (\ref{eq:Nalpha-scaling}) 
provide the local growth exponent $\alpha$ and the local scaling density $f(\alpha)$ up to 
additive constants $\sim 1/\ln L$. 
This is the reason why the curves shown in the Figs.~\ref{fig:mult1}, \ref{fig:mult2},
instead of being tangent to the line $y=\alpha$, intersect it.
The multifractal spectra presented in Refs.~\onlinecite{meakin:87:1,meakin:86:1,schwarzer:92:0}  
were also calculated as numerical histograms, and display the same behavior. 

\begin{figure}[]
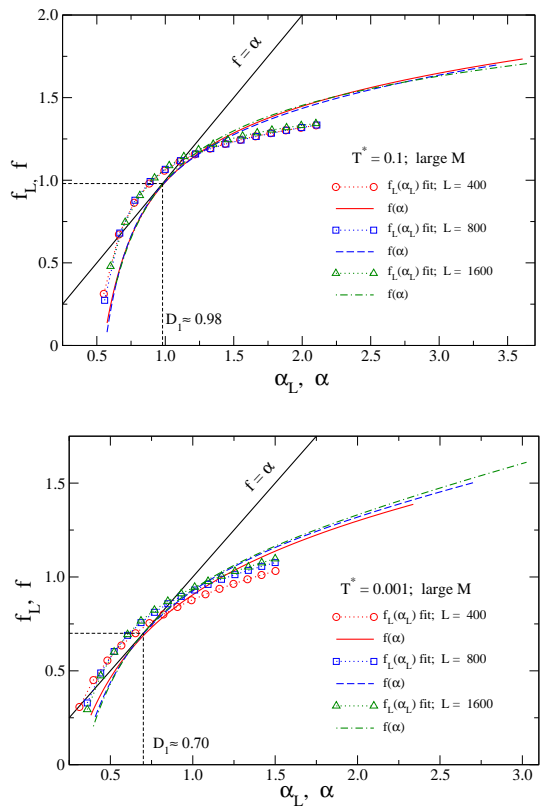

\begin{center}
\includegraphics[width=7cm]{figures/0.1_largeM_corr.eps}
\vspace*{0.5cm}

\includegraphics[width=7cm]{figures/0.001_largeM_corr.eps}
\end{center}
\caption{Corrected spectra $f(\alpha)$ calculated according to the 
Eqs.~(\ref{eq:alpha_assympt}),(\ref{eq:f_assympt}) for $T^*=0.1$, upper panel, and $T^*=0.001$, lower panel.
The numerical curves $f_L(\alpha_L)$ are fitted to the function $a_0 + a_1\exp(a_2\alpha_L^{a_3})+a_4\ln(\alpha_L)$.
The symbols represents the fitted functions, and the lines -- corrected spectra $f(\alpha)$.
For $T^*=0.1$ the numerical data  only in the range $\alpha_{Lmin}\leq \alpha_L \leq 2.1$, and for
$T^*=0.001$ in $\alpha_{Lmin}\leq \alpha_L \leq 1.5$, are fitted.}
\label{fig:corrected}
\end{figure}

From a more physical point of view, $f(\alpha)=\alpha$ is a 
turning point that separates regions of the spectrum corresponding to high (small $\alpha$) and low (large $\alpha$) 
growth probabilities. Moreover, the subset satisfying $f(\alpha)=\alpha$ 
has a fractal dimension equal to the information dimension $D_1$ of the multifractal formalism introduced by Hentschel 
and Procaccia. \cite{hp} The particles in this subset carry almost all the growth probability, in such a way that $D_1$ is 
the fractal dimension 
of the active zone, i.e. the unscreened surface. 
The identification of $D_1$ from Figs.~\ref{fig:mult1}, \ref{fig:mult2} is troublesome 
since there is no tangency at $f_L=\alpha_L$. 
The use of the corrected expressions (\ref{eq:alpha_assympt}) and (\ref{eq:f_assympt}) is therefore called for,
but their application immediately poses the problem of computing up to third order numerical derivatives from
noisy, scarce simulation data. We therefore resorted to fitting to the simulation data several functional forms.
In Fig.~\ref{fig:corrected} results for the corrected $f(\alpha)$ are shown for the particular choice of the
fitting function $a_0 + a_1\exp(a_2\alpha_L^{a_3})+a_4\ln(\alpha_L)$. 
 The numerical data are  fitted only for the values of $\alpha_L$ in the range
$\alpha_{Lmin}\leq \alpha_L \leq 2.1$ for $T^*=0.1$, and $\alpha_{Lmin}\leq \alpha_L \leq 1.5$
for $T^*=0.001$. For the larger values of $\alpha_L$ there is a big uncertainty in the 
numerical data, thus those parts of the $f_L(\alpha_L)$ curves are excluded from this particular fitting
procedure. We stress that the Fig.~\ref{fig:corrected} is shown only as a qualitative illustration
of the effects of the corrections terms  in equations (\ref{eq:alpha_assympt}) and (\ref{eq:f_assympt}).
In particular, the tangency property is restored in every case, while data
collapse is preserved. Estimating $D_1$ as the value of $f$ where the $f$-curve has unit slope we obtain 
$D_1\approx 0.98$ for $T=0.1$, which
is close to the exact result $D_1=1$ 
for DLA. \cite{makarov} With decreasing  $T^*$
we observe the decrease of $D_1$, namely we find $D_1\approx 0.84$ and $D_1\approx 0.70$ for $T^*=0.01$ 
(the corresponding figure is not shown) and $T^*=0.001$ respectively. Other functional forms yield very similar results.

The corrected spectra have other interesting features as compared to the uncorrected ones.
At this point, however, the discussion has to be kept at a qualitative level since at the extreme
values of $\alpha_L$ one faces two different numerical problems: 
i) on approaching $\alpha_{min}$ the
derivatives of any order of ``true'' spectra $f$ diverge; ii) on the other hand, large $\alpha_L$  values
correspond to data with poor statistics. Some general observations can nevertheless be made.     
There is strong indication that, upon incorporation of the corrections terms:
i) the part of the $f_L$-curve corresponding to large $\alpha_L$ 
shifts upwards for every temperature; ii) data collapse emerges for $T^* = 0.001$.
 In particular, for the values of 
$\alpha \approx 3.0$, $f(\alpha)$ takes similar values for both values of $T^*$, a
result consistent with the interpretation that the fractal dimension does not depend on the effective temperature.
Note also that a remarkable improvement in data-collapse can be observed for the case 
$T^*=0.001$ and large $M$. Finally, there is a slight tendency for $\alpha_{min}$ to
move to the right. For $T^*=0.1$ the corrected value gets even closer to its DLA value whereas
for $T^*=0.001$ it keeps at $\alpha_{min}\approx 0.4$. We stress that on the whole this same behavior  
carries over to other fitting choices.
We  also calculate the corrections to the measured $f_L(\alpha_L)$ spectrum in the
dipolar scaling regime $(T^*=0.001)$ and small values of $M)$. 
As expected we get the same $\alpha_{min}$ and $D_1$ as for large $M$, but
no concluding evidence that $D$ is going to be the same.

According to the Turkevich-Scher conjecture
 $D=1+\alpha_{min}$, \cite{turkevich:85:0} such behavior of $\alpha_{min}$
 would be in contradiction with the fact that $D$
 does not change with effective temperature (as concluded in
 section \ref{subsec:paco}).
 The above mentioned relation between $D$ and $\alpha_{min}$ is obtained
 using implicitly the assumption
 that the most extremal sites of the clusters, or tips, are the most active
 ones, i.e. $\alpha_{tip}=\alpha_{min}$. We have checked the
 location
 of the sites with maximum growth probability in the clusters of
 simulations, and found that they often do not
 coincide with the clusters tips. Thus, $\alpha_{tip}\geq\alpha_{min}$
and
 the more general relation $D\geq 1+\alpha_{min}$, Ref.~\onlinecite{halsey:book}, applies in this case.

\section{Conclusions}
\label{sec:conclusion}

The diffusion-limited deposition of magnetic particles shows a crossover
between two regimes, with a crossover size that sensitively depends on the temperature.
At the early stages of growth both the fractal dimension of the trees $D_t$
and their size-distribution, as given by the exponent $\tau$,  are temperature
dependent and have significantly smaller values as compared with those of pure
DLD (dipolar regime).
As the size of the trees exceeds the crossover value $s^*$ the diffusion-driven
DLA scaling regime emerges. In this regime $D_t$ and $\tau$ have the same values as
in DLD. \cite{tasinkevych:04:1}
Here, we have provided evidences that the fractal dimension of the entire deposit
remains constant as the deposition proceeds. This has been done by analyzing
the density profile of the deposit, the lateral correlation function,
and the mean height and the width of the upper surface.
It ensues that $D_t$ and $\tau$ conspire so as to give a fixed value of $D$.

Multifractal analysis shows that, for each value of the interaction parameter
and at each stage of growth, the normalized distribution of growth probabilities
can be scaled onto a single curve using the same scaling form as in DLD
thereby providing the $f(\alpha)$-spectrum.
The features of the $f(\alpha)$  spectra allows us to
distinguish the structure of the deposits at high and low temperature
(see Fig.~\ref{two_deposits}). We have
found that $\alpha_{min}$ decreases significantly with decreasing temperature,
revealing that
the concentration of the growth probability becomes more and
more marked when dipolar interactions are increased.
Likewise, the fractal dimension of the active zone $D_1$
decreases with decreasing temperature, meaning that
for low temperatures less sites are involved in the growth of the deposit.
As a consequence, and since $D_1$ and $\alpha_{min}$ do not depend on the stage of growth,
the presence of dipolar interactions reveals in the structure of
the deposits by increasing further the probability of growth at ``hotter'' sites
and by originating less dense deposits.
On the other hand, the $f(\alpha)$ spectra at late stages of growth
suggest that the fractal dimension of the deposits
does not depend on temperature (or dipolar interaction),
corroborating our previous results.
However, our findings also indicate that the $f(\alpha)$-spectra in the high- and low-temperature
DLA scaling regimes are different.

The multifractal spectrum in the dipolar regime (low temperature
at the early stages of growth) is clearly different from that in the initial
stages of growth at high temperature, thus providing evidence that these two situations
have to be held distinct.
The $f(\alpha)$ spectrum obtained for the dipolar regime was, however, not
accurate enough to confirm that the fractal dimension of the deposits
in this regime is approximately equal to that of later stages of growth.

In a more general perspective, this work shows that the information
dimension $D_1$ and the scaling exponent $\alpha_{min}$ are much easier to
determine through a numerical measurement of the $f(\alpha)$ spectrum then the
fractal dimension $D$, provided that the corrections of
Eqs.~(\ref{eq:alpha_assympt}) and (\ref{eq:f_assympt})  are taken into account. In fact, the low $\alpha$ part of
a spectrum can be obtained with good statistics  using small
systems (see Figs.~\ref{fig:mult1} and \ref{fig:mult2}).  Therefore, our work suggests that
the effect of interactions in DLA (or DLD) deposits could be more easily
studied through $D_1$ and $\alpha_{min}$.

\acknowledgments
We acknowledge M.M. Telo da Gama for critical reading of the
manuscript. M.T. would like to thank  M.N. Popescu and S. Kondrat for fruitful discussions.


\begin{thebibliography}{99}



\bibitem{witten_sander:81}
T. A. Witten and L. M. Sander,
Phys. Rev. Lett. {\bf 47}, 1400 (1981).


\bibitem{meakinbook}
P. Meakin, {\it Fractals, scaling and growth far from equilibrium
}, Cambridge University Press, Cambridge (1998).

\bibitem{meakin:83:2}
P. Meakin, Phys. Rev. A, {\bf 27}, 2616 (1983).

\bibitem{meakin:85:1}
P. Meakin, H. E. Stanley, A. Coniglio, and T. A. Witten 
Phys. Rev. A {\bf 32}, 2364 (1985).

\bibitem{halsey:86:1}
T. C. Halsey, P. Meakin, and I. Procaccia,
Phys. Rev. Lett. {\bf 56}, 854 (1986).


\bibitem{meakin:83:1}
P. Meakin,
J. Chem. Phys. {\bf 79}, 2426 (1983).

\bibitem{meakin:89:1}
P. Meakin and M. Muthukumar,
J. Chem. Phys. {\bf 91}, 3212 (1989).

\bibitem{vandewalle:95:1}
N. Vandewalle and M. Ausloos, Phys. Rev E {\bf 51}, 597 (1995). 

\bibitem{somfai:03:0}
E. Somfai, R. C. Ball,  N. E. Bowler, and L. M. Sander,
Physica A {\bf 325}, 19 (2003).

\bibitem{rubis}
R. Pastor-Satorras and J.M. Rub\'{\i},
Phys. Rev. E {\bf 51}, 5994 (1995);
Prog. Colloid. Polym. Sci. {\bf 110}, 29 (1998);
J. Magn. Magn. Mater. {\bf 221}, 124 (2000).



\bibitem{mors:87:1}
P.M. Mors, R. Botet, and R. Jullien,
J. Phys. A {\bf 20}, L975 (1987).

\bibitem{helgesen:88:1}
G. Helgesen, A.T. Skjeltorp, P.M. Mors, R. Botet, and R. Jullien,
Phys. Rev. Lett. {\bf 61}, 1736 (1988).

\bibitem{tasinkevych:04:1}
F. de los Santos, J. M. Tavares, M. Tasinkevych, and M. M. Telo da Gama,                                    
 Phys. Rev. E {\bf 69},  061406  (2004).   

\bibitem{meakin:86:3}
P. Meakin and F. Family, Phys. Rev. A {\bf 34}, 2558 (1986).

\bibitem{racz:83:0}
Z. R\'acz and T. Vicsek,
Phys. Rev. Lett. {\bf 51}, 2382 (1983).


\bibitem{stanley:77:0}
H. E. Stanley,
J. Phys. A: Math. Gen. {\bf 10}, L211 (1977).

\bibitem{schwarzer:92:0}
S. Schwarzer, M. Wolf,  S. Havlin, P. Meakin, and 
H. E. Stanley,
Phys. Rev. A {\bf 46}, R3016 (1992).
 



\bibitem{tasinkevych:03:1}
F. de los Santos, M. Tasinkevych, J.M. Tavares, and P.I.C. Teixeira,
J. Phys.: Condens. Matter {\bf 15}, S1291 (2003).
 
\bibitem{tavares:02:1} 
J. M. Tavares, J. J. Weis, and M. M. Telo da Gama,
Phys. Rev. E (2002).

%

%
%
%
%
%

\bibitem{meakin:84}
P. Meakin, Phys. Rev. B {\bf 30}, 4207 (1984).

\bibitem{differences}
Notice that in \cite{rubis}:
i) the dynamics of the dipoles is different: the orientation of their dipolar moments does not
change between two consecutive steps, but after each accepted step 
it relaxes along the total field at the arrival lattice site;
ii) deposition occurs on a single growth site rather than on a line (slab geometry);
iii) 2d dipoles satisfying 3d electrostatics are used.

\bibitem{ball:90:0}
R.C. Ball and O.R. Spivack,
J. Phys. A: Math. Gen. {\bf 23}, 5295 (1990).



\bibitem{amitrano:86}
C. Amitrano, A. Coniglio, and F. di Liberto,
Phys. Rev. Lett. {\bf 57}, 1016 (1986).
%
%
\bibitem{hayakawa:87}
Y. Hayakawa, S. Sato, and M. Matsushita,
Phys. Rev. A {\bf 36}, 1963 (1987).

\bibitem{meakin:87:1}
P. Meakin,
Phys. Rev. A {\bf 35}, 2234 (1987).

\bibitem{hp}
H .G. E. Hentschel and I. Procaccia, 
Physica {\bf 8}D, 435 (1983). 

\bibitem{halsey:book}
T.~C.~Halsey in {\it Fractals' Physical Origin and Properties},
edited by L.~Pietronero (Plenum Press, New York, 1989), p.205.


\bibitem{halsey:88:0}
T. C. Halsey, Phys. Rev. A {\bf 38 }, 4789 (1988).


\bibitem{meakin:86:1}
P. Meakin,
Phys. Rev. A {\bf 34}, 710 (1986).  

\bibitem{makarov}
N. G. Makarov,
Proc. Lon. Math. Soc. {\bf 51}, 369 (1985). 

\bibitem{turkevich:85:0}
L. Turkevich and H. Scher,
Phys. Rev. Lett. {\bf 55}, 1026 (1985); 
Phys. Rev. A {\bf 33}, 786 (1986).














\end{thebibliography}
\end{document}